\documentclass[a4paper,amsmath,amssymb,floatfix,prb,footinbib,twocolumn, preprintnumbers,showpacs]{revtex4}

\usepackage{graphicx}    
\usepackage{dcolumn}     
\usepackage{bm}          
\usepackage{color}
\usepackage[colorlinks=true ,citecolor=blue]{hyperref}%
\usepackage{dsfont}

\date{\today}

\begin{document}

\title{Proposal for an on-demand source of polarized electrons into the edges
   of a topological insulator}

\author{Andreas Inhofer}
\altaffiliation[Present address: ]{Laboratoire Pierre Aigrain, Ecole Normale Sup\'erieure, CNRS (UMR 8551), Universit\'e P. et M. Curie,
Universit\'e D. Diderot, 24, rue Lhomond, F-75231 Paris Cedex 05, France}
\affiliation{Freiburg Institute for Advanced Studies, Albert-Ludwigs-Universit\"at, D-79104 Freiburg, Germany}

\author{Dario Bercioux}
\affiliation{Freiburg Institute for Advanced Studies, Albert-Ludwigs-Universit\"at, D-79104 Freiburg, Germany}

\affiliation{Dahlem Center for Complex Quantum Systems and Institut f\"ur Theoretische Physik, Freie Universit\"at Berlin, Arnimallee 14, 14195 Berlin, Germany}

\begin{abstract}
We propose a device that allows for the emission of pairs of spin-polarized electrons into the edge-states of a two dimensional topological insulator. Charge and spin emission is achieved using a periodically driven quantum dot weakly coupled to the edge states of the host topological insulator. We present calculations of the emitted time-dependent charge and spin currents of such a dynamical scatterer using the Floquet scattering matrix approach. Experimental signatures of spin-polarized two-particle emission can be found in noise measurements. Here a new form of noise suppression, named $\mathbb{Z}_2$--antibunching, is introduced. Additionally, we propose a set-up in which entanglement of the emitted electrons is generated. This   entanglement is based on a post-selection procedure and becomes manifest in a violation of a Clauser-Horne-Shimony-Holt inequality.
\end{abstract}



\maketitle

\section{Introduction} 
Recent experiments have shown the possibility to emit quantized charges by mesoscopic capacitors~\cite{Gabelli:2006,feve:2007,Parmentier:2012}|these systems are also known as single-particle sources (SPSs). The indistinguishable nature of the emitted charges has been proven via a quantum interference experiment.\cite{Bocquillon:2013} A mesoscopic capacitor consists of a quantum dot (QD) that can exchange carriers only via a quantum point contact (QPC) with an external electron reservoir|usually a two-dimensional electron gas (2DEG). The carrier emission from the QD to the 2DEG is obtained by driving the QD's energy levels periodically around the Fermi level of the 2DEG via a gate coupled capacitively to it.~\cite{Moskalets:2008,Parmentier:2012} 
A perpendicular magnetic field is applied to the sample, so to be in the regime of the integer quantum Hall effect (IQHE). Thus, the system works by using the edge channels of the IQHE.

Since 2005, a new class of materials known as topological insulators (TIs) have attracted the attention of the physics community because they show conductive edge-- states, even in absence of external magnetic fields.\cite{Hasan:2010} Two-dimensional (2D) TIs have been theoretically predicted~\cite{Bernevig:2006} and experimentally verified~\cite{Koenig:2007}  first in HgTe/CdHgTe quantum wells at cryostat temperature of $4.2$~K. However, these systems did not show a perfect quantization of the edge states|this problem has been recently addressed in terms of charge puddles.~\cite{vayrynen:2013} The conductive edge--states  have a linear gapless energy dispersion. These can be described by a Dirac-like Hamiltonian where the chemical potential can be shifted by opportune gating of the system, similarly to graphene.\cite{Koenig:2007,Bernevig:2006} The edge nature of the particle transport in this material has  been further addressed by specific experiments investigating the current distribution along the transversal direction of the sample.~\cite{Koenig:2013,novack:2013} Nowadays, a race for finding new 2D TI is running, one of the most promising candidates are InAs/GaSb quantum wells.~\cite{liu:2008,knez:2011} In this type of heterostructure quantum wells the edge state quantization is more accurate. Further, they present a protection against time-reversal symmetry (TRS) perturbations, \emph{e.g.} external magnetic fields.\cite{du:2013}  In 2D TI systems, charge carriers with opposite spins move dissipationless in opposite directions on a given edge.\cite{Koenig:2007,Koenig:2013,novack:2013,Buettner:2011,Kane:2005} 

In this Article, we  extend the concept of the SPS beyond the IQHE paradigm. Contrary to the SPS, we plan to use the edge states of a 2D  TI, thus without external magnetic fields. The device we propose emits two charge particles at the same time. However, our device differs from a double SPS~\cite{Splettstoesser:2008,Bocquillon:2013} because the two emitted particles are not only  correlated in time but also in spin: their reciprocal spin polarization is fixed by TRS, thus forming a Kramers' pair. Because of these properties we name our device: Spin Particles Source (SpPS).

We believe the SpPS can represent the analog of the spin polarized photon source used in  quantum optics.~\cite{Aspect:1982} In this respect, we propose a scheme for the entanglement of the two emitted particles by virtue of a post-selection procedure. This paves the way to test the Clauser-Horne-Shimony-Holt~\cite{Clauser:1969} (CHSH) inequality in condensed matter physics complementing the various schemes already presented for the IQHE set-ups\cite{Beenakker:2003,Samuelsson:2004,Beenakker:2006} and Cooper pair splitters.~\cite{martin:1996,recher:2001,Hofstadter:2009,hermann:2010,Hofstadter:2011} Therefore we are confident that the SpPS could represent a novel approach to \emph{electron quantum optics}.
Further, we believe that the SpPS can also be an important building block of spintronics.\cite{zutick:2004} It can be used for creating on-demand pure spin-current.\cite{scheid:2007,smirnov:2008,governale:2003,citro:2011,Dolcetto:2013}

The Article is structured in the following way: we start by presenting the SpPS set-up (Sec.~\ref{sec:II}), then we show the results for the emitted current and noise (Secs.~\ref{sec:III} and \ref{sec:IV}, respectively). Finally,  we propose a scheme for entangling the Kramers' partners emitted by the SpPS and we propose a scheme for violating the CHSH inequality|that is known as a reformulation of the Bell inequality (Sec.~\ref{sec:V}). The Article concludes with some final remarks and a discussion about the experimental feasibility of the SpPS. Many technical details are summarized in the conclusive appendices.
%
%
\begin{figure}[!b]
	\centering
	\includegraphics[width=0.7 \columnwidth]{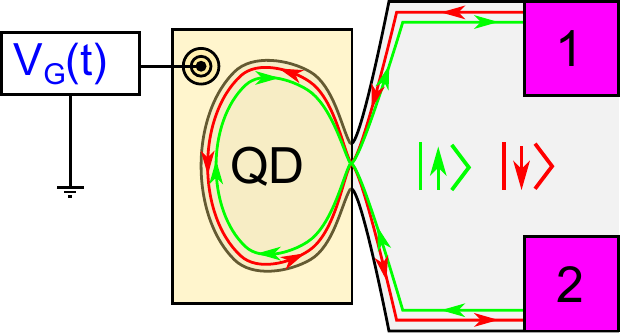}
	\caption{\label{setup} (Color online) Sketch of the Spin Particles Source. The quantum dot is supposed to be of circular shape with a perimeter of length $L$. The Kramers' partners are indicated by the ket states $|\!\uparrow\rangle$ and $|\!\downarrow\rangle$ (See main text for further details).}
\end{figure}
%
%


\section{Model and Formalism\label{sec:II}}

The system, represented in Fig.~\ref{setup}, is constituted by a QD shaped in a 2D TI structure, \emph{e.g.} by etching or opportune gating of a HgTe/CdHgTe or InAs/GaSb quantum well. A QPC separates the QD and the part that is connected to macroscopic electrodes, named the \emph{external part}. Furthermore, a metallic gate is placed on top of the QD.

We start by considering the edge states of the 2D TI. In the low energy approximation, these  can be described by an effective Hamiltonian, characterizing the spin polarized fermionic modes with linear dispersion:\cite{Hou:2009}
%
%
\begin{align}\label{eq:zero}
\mathcal{H}_0 =-\text{i} \hbar v_\text{F} &\sum_{\sigma,\bar{\sigma}=\uparrow,\downarrow} \int dx \left[ :\Psi^\dag_{\text{R}\sigma}(x)\partial_x\Psi_{\text{R}\sigma}(x):\right.\nonumber\\
&\hspace{2cm}- \left.:\Psi^\dag_{\text{L}\bar{\sigma}}(x)\partial_x\Psi_{\text{L}\bar{\sigma}}(x):\right]\,.
\end{align}
%
%
Here $x$ denotes the longitudinal coordinate along the edge,  $v_\text{F}$ is the Fermi velocity of the edge states, $\Psi_{\text{R},\sigma}$ and $\Psi_{\text{L},\sigma}$ are the right and left mover electron fields, respectively. The index $\sigma =\uparrow ,\downarrow$ is the spin component with respect to the a generic quantization axis. In fact, we neglect the Rashba spin-orbit interaction between the valence and the conductance band, thus we consider constant the spin projection of the two edge channels.~\cite{rothe:2010} In this representation, the spin eigenstates $\sigma$ and $\bar{\sigma}$ form Kramers' pairs. At equilibrium, we assume the Fermi level to be at the Dirac point of the spectrum.
The symbol $:\, :$ in Eq.~\eqref{eq:zero} denotes normal ordering with respect to the ground state density.

The QD is characterized by a set of discrete levels corresponding to quantized counter propagating edge states|the quantization is related to the perimeter $L$ of the QD itself. The time for a complete round trip along the edge of the QD is $\tau=L/v_\text{F}$. Thus, the QD level spacing is  $\Delta=h/\tau=h v_\text{F}/L$.

The metallic gate couples capacitively to the QD, this allows for shifting its energy levels periodically in time:
%
%
\begin{equation}\label{eq:gate}
H_\text{g}(t) = \int_{0}^{L} dy\ e V_\text{g}(t) \left[ \rho_{\text{R}\uparrow}(y) + \rho_{\text{L}\downarrow}(y) \right]
\end{equation}
%
%
where $\rho_{\alpha\sigma}(y)=:\!\Psi_{\alpha\sigma}^\dag(y)\Psi_{\alpha\sigma}(y)\!:$ is the electron density operator, and $y$ the coordinate along the QD perimeter. The gate potential $V_\text{g}(t)=V_0+V_\text{ext}(t)$ contains  two terms: a time independent one $V_0$, used for tuning the position of the energy levels of the QD with respect to the Fermi energy of the external part; and a periodic time-dependent part $V_\text{ext}(t)$ that shifts these energy levels in time. 

Experimental data for the case of the SPSs~\cite{Gabelli:2006,feve:2007,Parmentier:2012,Bocquillon:2013} has shown that Coulomb interaction for dots of size $\sim 1\mu$m in diameter has as main effect the renormalization of the Fermi-velocity, but does not affect qualitatively the particle emission. Recent experiments have shown that the edge states are characterizing by a width on the order of 5 $\mu$m.~\cite{Koenig:2013,novack:2013} In order to avoid overlap between the edge states, we estimate that the radius of the SpPS QD should be greater than the experimental measured  width of the edge states. Within this assumption the screening of the Coulomb interaction induced by the top gate enables us to treat the system as non-interacting. Thus, we proceed using a Landauer-B\"uttiker treatment as underpinned by Ref.~[\onlinecite{Teo:2009}].

We start by analyzing the QPC. Due to TRS, only three processes at a QPC are allowed, namely: (1) a spin-preserving forward scattering, (2) a spin- and edge-flipping forward scattering and (3) a spin-preserving edge-flipping backscattering. The relevant Hamiltonians at the QPC were already investigated in previous works~\cite{Teo:2009,Dolcini:2011,Krueckl:2011} [for clarity we recall them in App.~\ref{app:one}].

The calculation of the energy-independent scattering matrix of a QPC in a 2D TI is straightforward\ [c.f. App.~\ref{app:one}] and reads:
%
%
\begin{equation}\label{QPS:SCM}
\Sigma=\begin{pmatrix}
0 & \lambda_\text{pb} & \lambda_\text{ff} & \lambda_\text{pf} \\ 
\lambda_\text{pb} & 0 & \lambda_\text{pf} & \lambda_\text{ff} \\ 
\lambda_\text{ff}^\ast & \lambda_\text{pf} & 0 & \lambda_\text{pb} \\ 
\lambda_\text{pf} & \lambda_\text{ff}^\ast & \lambda_\text{pb} & 0
\end{pmatrix} 
\end{equation}
%
%
where $\lambda_\text{pb}$, $\lambda_\text{ff}$, $\lambda_\text{pf}$ are the scattering amplitudes for the spin-preserving back-scattering, the spin-flipping forward-scattering and the spin-preserving forward-scattering, respectively.
The structure of this scattering matrix is imposed by TRS. A simple Hamiltonian model allows to parametrize the elements of the scattering matrix (c.f.~App.~\ref{app:one} and Ref.~[\onlinecite{Dolcini:2011}]): 
%
%
\begin{equation*}
\lambda_\text{pb}=  -2\text{i}  \frac{\gamma_\text{sp}}{1+\gamma_\text{sf}^2 + \gamma_\text{sp}^2}, \lambda_\text{ff}= 2\text{i} \frac{\gamma_\text{sf}}{1+\gamma_\text{sf}^2 + \gamma_\text{sp}^2},
\end{equation*}
%
\begin{equation*}
\lambda_\text{pf}=\frac{1-\gamma_\text{sf}^2-\gamma_\text{sp}^2}{1+\gamma_\text{sf}^2 + \gamma_\text{sp}^2},
\end{equation*} 
%
%
where $\gamma_\text{sf}$ and $\gamma_\text{sp}$ are dimensionless amplitudes for spin-flipping processes and spin-preserving backscattering processes, respectively.

Next, we evaluate the time dependent two-terminal scattering matrix of the QD|Fig.~\ref{setup}. Therefore the wave-functions turning clockwise (spin-up) and anti-clockwise (spin-down) inside the QD are matched to the edge-state wave-function of the external region. 
The additional time-dependent gate-voltage is accounted for by  using the Floquet scattering matrix (FSM)\ (c.f.~App.~\ref{app:two} and Ref.~[\onlinecite{Moskalets:2011}]; an alternative approach is based on a time-dependent tight-binding approach in~Ref.~[\onlinecite{Jonckheere:2012}]).
The Fourier-transform of the FSM is a time-dependent $4\times4$ matrix $\mathcal{S}(t,E)$ corresponding to the scattering amplitude for a particle incident on the time-dependent scatterer with energy $E$ and leaving the scattering region at a time $t$. Due to the distinctive spin-momentum locking of 2D TI, the S-matrix reduces to a $2\times2$ form. Furthermore,
conservation of TRS implies: $\mathcal{S}_{11}=\mathcal{S}_{22}=0$ and $\mathcal{S}_{12}=\mathcal{S}_{21}$. The off-diagonal elements read:
%
%
\begin{align}\label{eq:scattmat:2}
\mathcal{S}_{12}(t,E)=\lambda_\text{pb} + &(|\lambda_\text{ff}|^2 + \lambda_\text{pf}^2)  \\
	&\nonumber\times\sum_{q=1}^\infty \lambda_\text{pb}^{q-1} \text{e}^{\text{i} q k_E L-\text{i} \phi_q(t)}\,, 
\end{align}
%
%
where  $k_E=E/(\hbar v_\text{F})$ is the momentum and 
%
%
\begin{equation}\label{eq:phase:factor}
\phi_q(t)=\frac{e}{\hbar} \int_{t-q\tau}^t V_\text{g}(t') dt'
\end{equation}
%
%
is a time-dependent phase factor due to the gate.

The matrix element $\mathcal{S}_{12}$ represents Fabry-P\'erot modes with an additional phase due to the time-dependent gate $\phi_q(t)$. The first term corresponds to direct reflection, the other parts to multiple tours clockwise or counterclockwise within the QD [c.f.~Fig.~\ref{figure:processes} in App.~\ref{app:two}]. It is interesting to notice, that|as reflected by the vanishing diagonal elements of the S-matrix and the helical nature of the edge-states|although a particle with spin up incident on the QD, can leave it only as a spin up, the spin state within the QD is not defined. The states inside the QD are in a superposition of spin-up and spin-down propagating clockwise or anti-clockwise, respectively.


\section{Emitted currents\label{sec:III}} 

Here we evaluate the charge and spin currents in terms of the time-dependent scattering matrix $\mathcal{S}(t,E)$. It is important to note that in each lead, spin and charge currents are not separated. An electron emitted by the SpPS towards lead one/two will always be a spin-up/down state [c.f.~App.~\ref{app:three}].  

\begin{figure}[!t]
\includegraphics[width=0.9 \columnwidth]{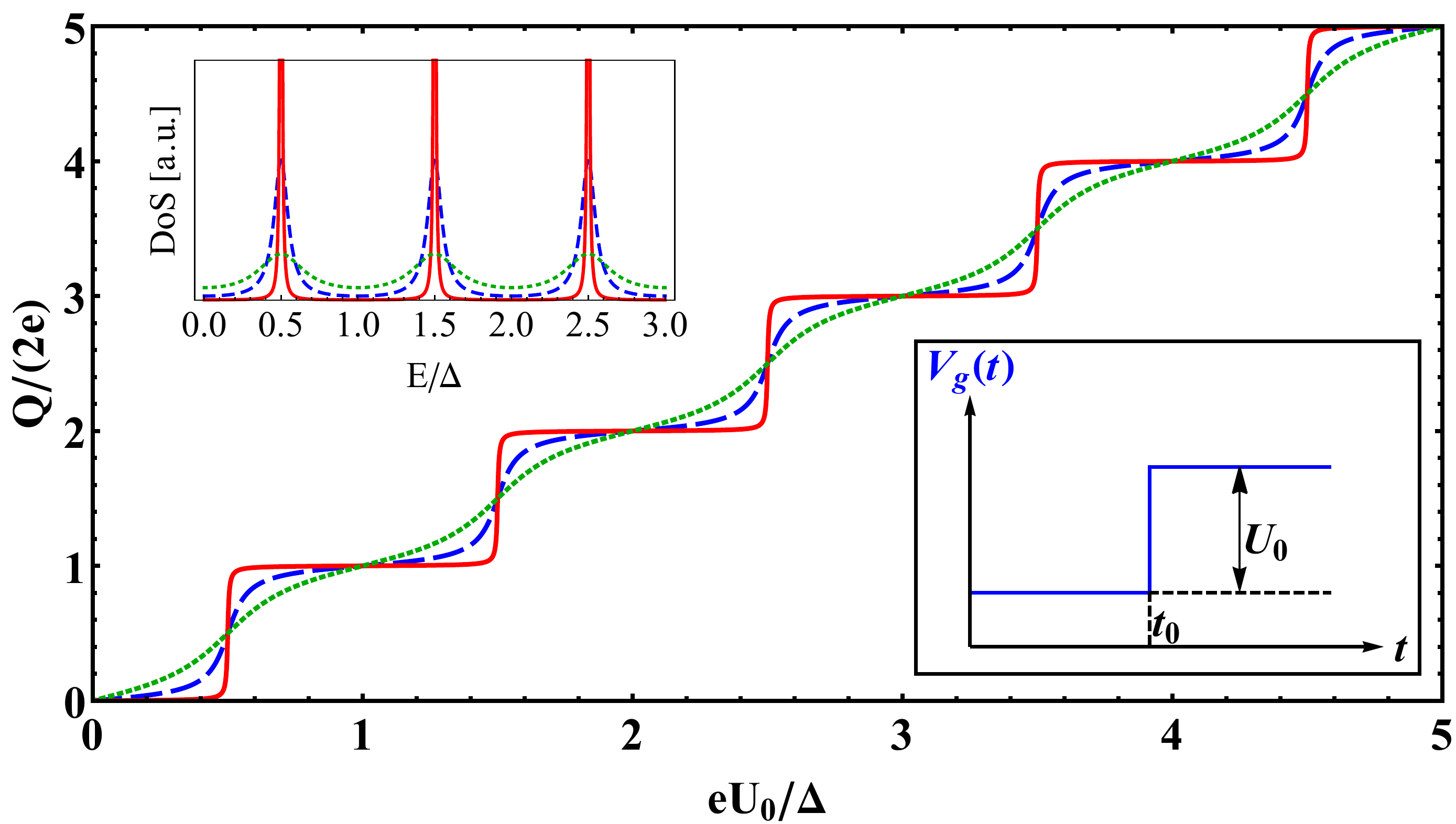}
\caption{\label{EmittedCharge} (Color online) Zero temperature emitted charge as a function of the height of the driving step $U_0$. The various lines correspond to different values of the reflection probability of the QPC $|\lambda_\text{pb}|^2$: $|\lambda_\text{pb}|^2=0.95$ (solid-red line), $|\lambda_\text{pb}|^2=0.5$ (dashed-blue line), and $|\lambda_\text{pb}|^2=0.1$ (dotetd-green line). Please note that the steps are given in units of twice the electron charge. (Upper Inset) The density of states of the QD for various values of $|\lambda_\text{pb}|$. (Lower Inset) Model of driving we consider: the driving of the  gate is switched on at time $t_0$ and the potential is changed by a quantity $U_0$.}
\end{figure}

Low-noise, single carrier emission of quantized spins/charges is obtained under the so called conditions of optimal driving. These are met if the static part of the gate potential $V_0$ shifts one QD level at resonance with the Fermi-energy of the external part and the peak-to-peak distance of the time-dependent potential equals exactly one level-spacing of the dot, c.f. Fig.~\ref{EmittedCharge}. The current in lead $\alpha$ can be expressed as a sum of two contributions:~\cite{Moskalets:2008} $I_\alpha(t)=I_\alpha^\text{d}+I_\alpha^\text{od}$, where the contribution $I_\alpha^\text{d}$ and $I_\alpha^\text{od}$ arise from the diagonal and off-diagonal elements in the double-sum of the product of scattering matrices, respectively. The diagonal contribution reads 
%
%
\begin{equation}\label{eq:curr:diag}
I_\alpha^\text{d}=\frac{e^2 (1-|\lambda_\text{pb}|^2)}{h}\sum_{q=1}^\infty |\lambda_\text{pb}|^{2q-2} \delta V_\text{g}(q\tau)\,, 
\end{equation}
%
%
where $ \delta V_\text{g}(q\tau)=\left[V_\text{g}(t)-V_\text{g}(t-q\tau)\right]$.
This element of the emitted current is temperature-independent and cannot yield any robust quantization of the emitted charge.~\cite{Moskalets:2011}

The off-diagonal elements in the double sum yield a temperature-dependent expression with a fast suppression at high temperatures
%
%
\begin{align}\label{eq:curr:off:diag}
I_\alpha^\text{od}=&\frac{e (1-|\lambda_\text{pb}|^2)^2}{\tau\pi}\text{Im}\left\{ \sum_{s=1}^\infty\left(\frac{\eta\left(s\frac{T}{T^*}\right)\left(\lambda_\text{pb}\text{e}^{\text{i} k_\text{F} L}\right)^s}{s}\right)\right. \nonumber \\
& \times \left.\sum_{q=1}^\infty  |\lambda_\text{pb}|^{2q-2}  \left( \text{e}^{-i\phi_s(t-q\tau)}-\text{e}^{-i\phi_s(t)}\right) \right\}\,,
\end{align}
%
%
where $\text{Im}[\ldots]$ stands for the imaginary part and the dependence on the effects of the gate potential are included in the phase factors \eqref{eq:phase:factor}. In Eq.~\eqref{eq:curr:off:diag}  we have introduced the Fermi momentum as $k_\text{F} = \mu/(\hbar v_\text{F})$ with $\mu$ the chemical potential of the external part, the function $\eta(x)=x/\sinh(x)$, and the crossover temperature $T^*=\Delta/(2\pi^2 k_\text{B})$, with $k_\text{B}$ the Boltzmann constant.~\cite{Moskalets:2008,Moskalets:2011}

We assume the experimentally relevant case of a square-shaped time-profile of the gate-voltage with a driving frequency smaller compared to all other relevant time scales. In this regime the emission of holes and particles can be treated independently.\cite{Moskalets:2013} Therefore it is enough to consider a step-potential [c.f. Lower inset of Fig.~\ref{EmittedCharge}]: 
%
%
\begin{equation*}
V_\text{ext}(t)=U_0\left[\frac{1}{2}-\theta(t_0-t)\right], 
\end{equation*}
%
%
where $\theta(x)$ is the Heaviside  function. 

In Fig.~\ref{EmittedCharge} the total emitted charge propagating towards the leads is plotted as a function of the height of the step $U_0$ for different values of the reflection probability of the QPC $|\lambda_\text{pb}|^2$. It shows  the quantization of the charge emitted to a single lead. For small transparencies of the QPC, the quantization on the plateaus is almost perfect. Only for higher transmission-probabilities (above $\approx 0.3$), the density of states for the dot is broadened, resulting in a blurring of the quantization. 
We deduce that the SpPS emits two fully time-correlated charge particles in a fashion very similar to a double SPS.\cite{Splettstoesser:2008,Bocquillon:2013} However, the two emitted particles have also opposite spin-polarization as imposed by TRS, forming a Kramers' doublet.
The corresponding emitted spin charge accumulated per lead is quantized in units of $\hbar/2$. Plugging a two-terminal spintronics device between the two contacts 1 and 2 in Fig.~\ref{setup} allows to operate the SpPS as a pure spin current generator. The total spin-current provided by the SpPS per half-cycle is then equal to $\hbar$ corresponding to two spin quanta.\cite{scheid:2007}

%
%
%
%


\section{Noise\label{sec:IV}} 
\begin{figure}[t]
\includegraphics[width=0.9 \columnwidth]{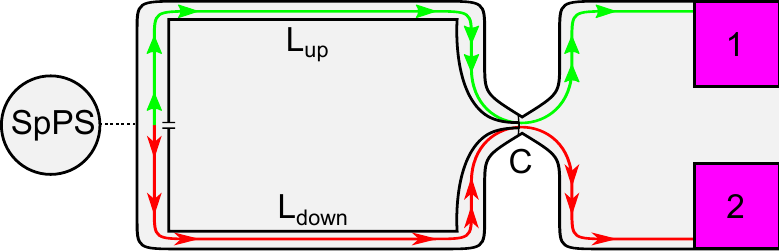}
\caption{\label{antibunching} (Color online) After the emission, the two spin-polarized electrons are separated spatially, propagating along the upper and the lower edge. A second QPC (C) allows for interference of the two particles. The QPC is tuned in a regime, where only forward scattering processes are allowed ($\lambda_\text{pb}=0$ for C). No reabsorption processes are considered.}
\end{figure}

Single electron-- or spin--detection at the experimental relevant frequencies of $\sim 1$GHz still remains a challenging task. However, signatures of the quantization can be found in noise measurements.\cite{Parmentier:2012,Blanter:2000} The autocorrelation function of the current measured in one of the leads allows for a characterization of one of the emission channels. We consider the correlation function between the two contacts  as defined in Ref.~[\onlinecite{Blanter:2000}]. At finite temperatures $T$, the only contribution to noise is thermal noise.
It does not contain any information about the quantization of the emitted charge.
This noise is independent of the QD and can therefore not provide any information about the SpPS. Interestingly the same result can be obtained in a slightly more complicated set-up|Fig.~\ref{antibunching}. Here, the two emitted particles are propagating towards a second QPC|called C|tuned in a regime where only forward scattering processes contribute. Such a regime can be obtained by changing the details of the QPC.\cite{Krueckl:2011} Then, higher order processes involving reabsorption of particles by the SpPS can be neglected. 
%
%
%
%
%
%
This set-up is an electronical solid state realization of the Hong-Ou-Mandel optics experiment.~\cite{hong:1987} Specifically, C is now acting as a beam splitter. If two electrons, emitted by the SpPS, are simultaneously incident onto C, two  scenarios are possible: (i) the two electrons are identical, \emph{i.e.} possess the same quantum numbers. In this case the Pauli principle gives rise to a phenomenon known as (electronic) anti-bunching. In this case, the electrons propagate towards different directions after scattering at C. This is in contrast to the case of the standard (optical) Hong-Ou-Mandel experiment in which photons tend to \emph{bunch}; (ii) the two electrons are not identical, \emph{i.e.} possess different quantum numbers. Then, after scattering at C, in lack of any quantum mechanical restriction, both electrons can move in the same direction. Experimentally, the former corresponds to an arrival at two different detectors and the latter to an arrival at the same detector.~\cite{Blanter:2000}

There has been a theoretical\cite{Moskalets:2013} and experimental analysis\cite{Bocquillon:2013} of the electronic anti-bunching for two synchronized SPS. This corresponds to the discussed former case where fermionic statistics lie at the heart of an anti-bunching effect. A distinctive feature of 2D TI is an effect, known as the $\mathbb{Z}_2$ peak of noise.~\cite{Edge:2013} It predicts an electronic anti-bunching in the case of two electrons with opposite spin, propagating in the edges of a 2D TI as long as TRS is conserved.
For the case of the SpPS, these two effects interplay and lead to a complete noise--canceling. This is somewhat surprising as one expects the two single-particle excitations arriving at C to occupy different spin states. Hence, the Pauli principle does not apply and thus the two particles could still bunch, however TRS conservation prevents them from moving along the same edge resulting therefore in an effective antibunch. Indeed, if they were propagating along the same edge, they would have to occupy the same spin state, what is ruled out by fermionic statistics. This is a new form of antibuching typical of TI, thus we name this phenomenon  $\mathbb{Z}_2$-\emph{antibunch}.

In our set-up the synchronization is included in the emission of the two particles. Thus, measuring zero correlations is a signature of synchronized emission of the two particles. Deviations from the zero-noise are introduced by a magnetic field applied locally to  the QD~\cite{hofer:2013,fnote:one}$^,$
or by changing the arrival time of the two electrons with respect to each other. The latter case can in principle be realized by changing the length of the upper $L_\text{up}$ or the lower $L_\text{down}$ arm in Fig.~\ref{antibunching} or equivalently by altering the Fermi-velocity in a finite region of one of the arms. This might be realized by application of an out-of-plane electrical field that introduces a Rashba spin-orbit interaction.\cite{Vayrynen:2011} In this case, the spin projections are not constant anymore, however, the two particles remain Kramers' partners with opposite spin polarization. By introducing such a time-delay $\delta \tau$, the diagonal elements of the scattering matrix become finite leading to a non-zero cross-correlator at zero temperature similar to the case studied in Ref.~[\onlinecite{Moskalets:2013}]. Assessment of the typical longitudinal width of the wave packet shows that they are rather broad (on the order of microns), corresponding to an uncertainty in the arrival time of the electrons at C. Hence, experimentally, one should expect a small deviation from the zero-noise feature, characteristic of perfect synchronization.


\section{Entanglement\label{sec:V}}
\begin{figure}
\includegraphics[width=0.9 \columnwidth]{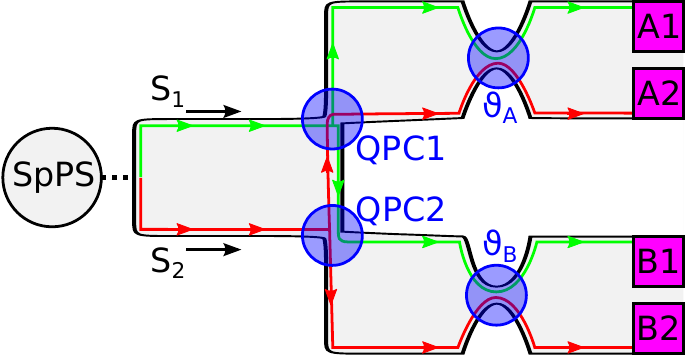}
\caption{\label{entangler} (Color online)~The SpPS creates two particles propagating along the edges of the 2D TI. Two QPCs allow for an exchange of these two particles, redistributing them to two parties ``Alice" and ``Bob". Post-selection after these two QPCs yields a non-vanishing concurrence. Two additional QPCs, the analogues of ``local" polarizers,  characterized by angles $\vartheta_\text{A/B}$ allowing for violation of the CHSH-inequality.}
\end{figure}

\begin{figure}
\includegraphics[width=0.9 \columnwidth]{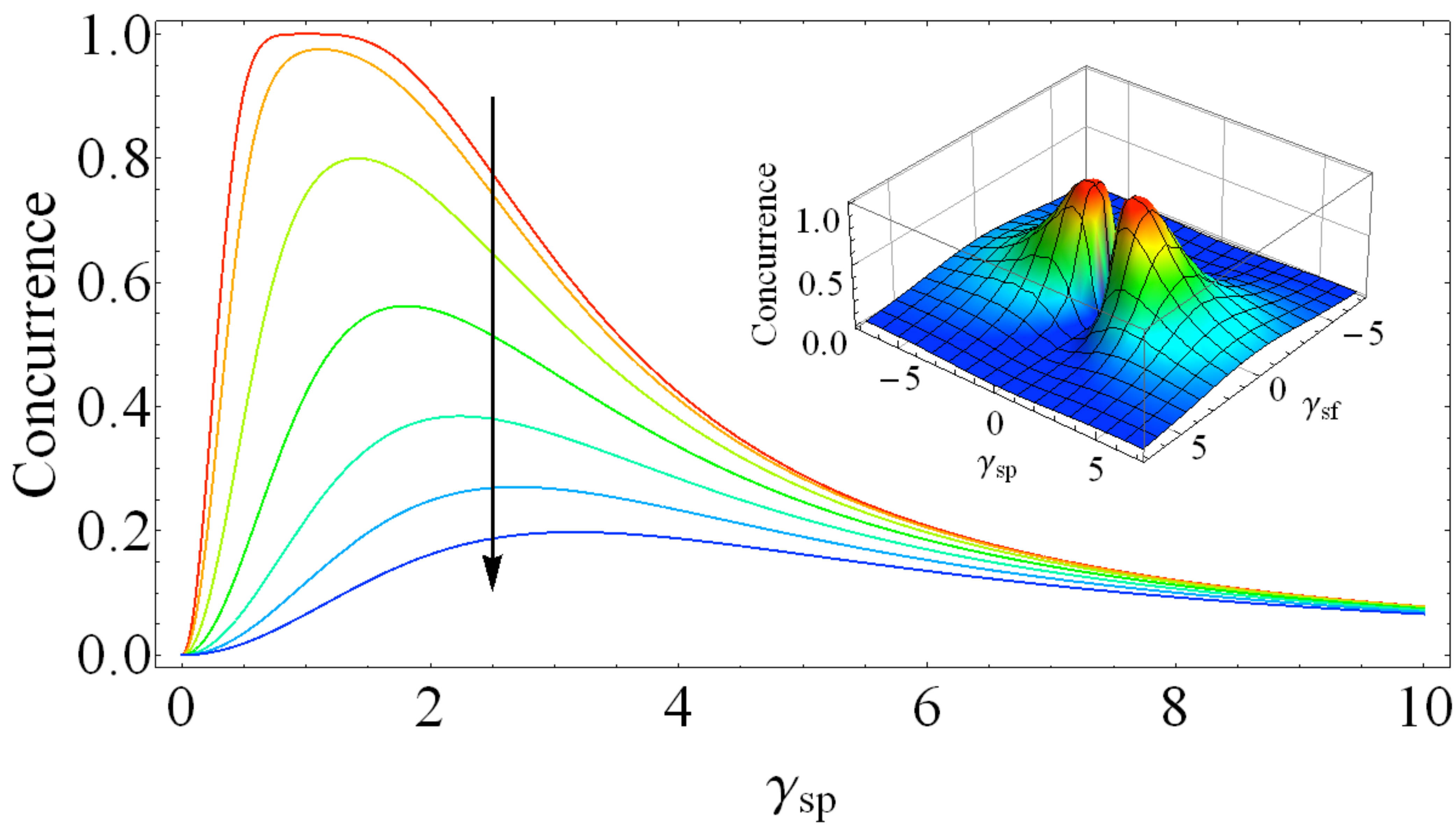}
\caption{\label{concurrence} (Color online) The concurrence after postselection of simultaneous detection events for symmetric QPCs as a function of the spin-preserving amplitude $\gamma_\text{sp}$ for different values of the spin-flipping amplitude $\gamma_\text{sf}$, ranging from $\gamma_\text{sf}=0$ to $\gamma_\text{sf}=3$ in steps of 0.5. The arrow indicates the direction of increasing $\gamma_\text{sf}$. (Inset) Concurrence as a function of $\gamma_\text{sf}$ and $\gamma_\text{pf}$.}
\end{figure}

The two particle state emitted by the SpPS and propagating to the right in Fig.~\ref{entangler}  is characterized in second quantization by
%
%
\begin{equation}\label{singlet}
|\Psi\rangle=a_{\text{S}_1}^\dagger a_{\text{S}_2}^\dagger |0\rangle,
\end{equation}
%
%
where $a_{\text{S}_{1/2}}^\dagger$ are creation operators for the SpPS-created wave-packets in the output arms S$_1$ (upper edge) and S$_2$ (lower edge).
%
%
%
%
This state is a Slater determinant and thus \textit{not} entangled.\cite{Schliemann:2001}
In the following we propose a set-up allowing for the entanglement of the two emitted particles. We consider bipartite entanglement between two parties where we adopt the standard notation of quantum information, calling the two parties Alice (A) and Bob (B). In the set-up, Fig.~\ref{entangler}, A (B) corresponds to the upper (lower) branch. The internal degree of freedom is a two-level system realized by the one-particle degree of freedom to propagate on the upper or lower edge of each party's output branch. Due to spin-momentum locking in 2D TI, this corresponds to spin qubits. The two particles are scattered by two additional QPCs, 1 and 2, realizing a 6-terminal set-up whose scattering matrix  reads
%
%
\begin{equation}
S^{(6)}=
\begin{pmatrix}
\mathcal{S}_\text{d}^{(1)} & \mathcal{S}_\text{od}\\
\mathcal{S}_\text{od}^\dag & \mathcal{S}_\text{d}^{(2)}
\end{pmatrix}
\end{equation}
%
%
this is a block matrix with elements defined by
%
%
\begin{equation}
\mathcal{S}_\text{d}^{(i)}=\begin{pmatrix}
 0 & \lambda_\text{pf}^{(i)} & \lambda_\text{ff}^{(i)}\\
\lambda_\text{pf}^{(i)} & 0 & \lambda_\text{pb}^{(i)} \\
\lambda_\text{ff}^{(i)^*} & \lambda_\text{pb}^{(i)} & 0
\end{pmatrix}
\end{equation}
%
%
and
%
%
\begin{equation}
\mathcal{S}_\text{od}=\begin{pmatrix}
\lambda_\text{pb}^{(1)} \lambda_\text{pf}^{(2)} & \lambda_\text{pb}^{(1)} \lambda_\text{ff}^{(2)\ast} & \lambda_\text{pb}^{(1)} \lambda_\text{pb}^{(2)} \\
\lambda_\text{ff}^{(1)\ast} \lambda_\text{pf}^{(2)} & \lambda_\text{ff}^{(1)} \lambda_\text{ff}^{(2)} & \lambda_\text{ff}^{(1)\ast} \lambda_\text{pb}^{(2)} \\
\lambda_\text{pf}^{(1)} \lambda_\text{pf}^{(2)} & \lambda_\text{ff}^{(2)\ast} \lambda_\text{pf}^{(1)} & \lambda_\text{pb}^{(2)} \lambda_\text{pf}^{(1)} 
\end{pmatrix},
\end{equation}
%
%
where the quantities $\lambda_i^{(\alpha)}$ describe exactly the same processes as before. The superscript $\alpha$ indicates the QPC, to which the quantity is assigned. In the following, we assume two symmetric QPCs, \emph{i.e.} $\lambda_i^\text{(1)}=\lambda_i^\text{(2)}$.
This allows to erase the which path information from the two emitted particles. On the output side (\emph{i.e.} right hand side in Fig.~\ref{entangler}), each particle lives in a four dimensional Hilbert space spanned by propagating modes in the output arms.  In this case, we can calculate the concurrence $\mathcal{C}$ of the state produced by the SpPS and subsequently scattered by the 6-terminal set-up composed of QPC1 and QPC2, using a formula derived in Ref.~[\onlinecite{Schliemann:2001}]: 
%
%
\[
\mathcal{C}=8\sqrt{\det{\tilde w}}.
\]
%
%
This gives a quantitative measure for the entanglement: $\mathcal{C}=1$ ($\mathcal{C}=0$) indicates maximum (no) entanglement. After scattering, the two qubit state is re-expressed as 
%
%
\begin{equation}\label{eq:33}
|\Psi\rangle=\sum_{i,j=1}^6 w_{i,j} b_{i}^\dagger b_{j}^\dagger |0\rangle
\end{equation}
%
%
with 
%
%
\begin{equation}\label{eq:w}
w_{i,j}=\frac{1}{2}	(S_{i,1}^{(6)}S_{j,6}^{(6)}-S_{i,6}^{(6)}S_{j,1}^{(6)}),
\end{equation}
%
%
where the $b^\dagger$-operators correspond to creation in one of the output arms.
Direct evaluation of the concurrence for the outgoing state yields zero, indicating a lack of entanglement. However, it has been shown \cite{Lebedev:2004} that a certain component of the wave-function addressed in a post-selection procedure, corresponds to a maximally entangled state. Therefore, the two-particle state is projected on the part of the wavefunction corresponding to simultaneous detection of particles by A and B. This post-selection restricts the coefficient matrix $w$ to the off-diagonal blocks. After this projection, the coefficients are renormalized in order to restore a unit modulus for the wave-function. Experimentally, this corresponds to choosing a subset of detection events. The emerging non-classical correlations are reflected in a non-vanishing concurrence as depicted in Fig.~\ref{concurrence}. We observe, that for a specific choice of the QPC-characteristic amplitudes $\gamma$, the concurrence is non-zero (equal to 1) indicating partial (maximum) entanglement.

For experimental detection of this entanglement, we propose a set-up that allows for violation of the CHSH-inequality.~\cite{Clauser:1969} Therefore, A and B have analogs of light polarizers at their disposal. These can be realized by two further QPCs, tuned in a regime, where only forward scattering is  allowed. These local transformations are parametrized in an S-matrix as
%
%
\begin{equation}
\begin{pmatrix}
b_{\text{A/B}_1}\\
b_{\text{A/B}_2}
\end{pmatrix}=\begin{pmatrix}
-\text{i} \cos(\vartheta_\text{A/B}) & \sin(\vartheta_\text{A/B}) \\ 
\sin(\vartheta_\text{A/B}) & -\text{i} \cos(\vartheta_\text{A/B})
\end{pmatrix} \cdot 
\begin{pmatrix}
b_{\text{A/B}_1}^\prime\\
b_{\text{A/B}_2}^\prime
\end{pmatrix}
\end{equation}
%
%
with the local ``polarization angles" $\vartheta_\text{A/B}$. Here, the primed operators are the annihilation operators for particles on the left of the local QPCs and the unprimed operators correspond to the states propagating to detectors A and B. Standard combination techniques of the different scattering matrices allow us to derive the two-particle state after local rotations and post-selection.
Such a tunability of transmission probabilities is predicted to be achievable using electrostatical side-gates on constrictions.\cite{Krueckl:2011} In the qubit picture, the effect of the two polarizer-QPCs corresponds to local rotations on the Bloch sphere.  
The CHSH-inequality reads
%
%
\begin{equation}\label{eq:CHSH}
|E(\vartheta_\text{A},\vartheta_\text{B})+E(\vartheta_\text{A}^\prime,\vartheta_\text{B})+E(\vartheta_\text{A},\vartheta_\text{B}^\prime)-E(\vartheta_\text{A}^\prime,\vartheta_\text{B}^\prime)|\leq 2\,.
\end{equation}
%
%
Here, the normalized particle number difference correlators are defined by
%
%
\begin{equation}\label{eq:correlators}
E(\vartheta_\text{A},\vartheta_\text{B})=\frac{\langle(\hat n_{\text{A}_1}-\hat n_{\text{A}_2})(\hat n_{\text{B}_1}-\hat n_{\text{B}_2})\rangle_{\vartheta_\text{A},\vartheta_\text{B}}}{\langle(\hat n_{A_1}+\hat n_{\text{A}_2})(\hat n_{\text{B}_1}+\hat n_{\text{B}_2})\rangle_{\vartheta_\text{A},\vartheta_\text{B}}}
\end{equation}
%
%
where $\langle \cdots \rangle_{\vartheta_\text{A},\vartheta_\text{B}}$ stands for averaging over the state produced by the four QPCs, provided the incident state is given by Eq.~\eqref{singlet} and Alice's and Bob's polarizers  characterized by the angles $\vartheta_\text{A}$ and $\vartheta_\text{B}$. In Eq.~\eqref{eq:correlators}, $\hat n_i=\hat b_i^\dagger \hat b_i$ with $(i\in \{$A$_1,\ldots,$B$_2\})$ is the electron number operator of output arm $i$. The envisioned detection scheme based on particle counting restricted to joint detection events implements the above-mentioned post-selection in a natural way. For symmetric QPCs and the parameters $\gamma_\text{sf}=0, \gamma_\text{sp}=1$, the CHSH-inequality takes a particularly simple form:
%
%
\begin{align}
|&\cos [2 (\vartheta_{\text{A}_1}+\vartheta_{\text{B}_1})]+\cos [2 (\vartheta_{\text{A}_1}+\vartheta_{\text{B}_2})]\\ &+\cos [2 (\vartheta_{\text{A}_2}+\vartheta_{\text{B}_1})]-\cos [2 (\vartheta_{\text{A}_2}+\vartheta_{\text{B}_2})]|\leq 2\nonumber
\end{align}
%
%
For the angles $\vartheta_{\text{A}_1}=0,\ \vartheta_{\text{A}_2}=3\pi/4,\ \vartheta_{\text{B}_1}=-3\pi/8,\ \vartheta_{\text{B}_2}=3\pi/8$ the left-hand side evaluates to $2\sqrt{2}$, thus maximally violating the inequality. As expected the maximum violation of the CHSH inequality corresponds to the maximum concurrence as shown in Fig.~\ref{concurrence}.
However, this result has to be handled with care. Strictly speaking, for $\gamma_\text{sf}=0, \gamma_\text{sp}=1$, the particles cannot reach A's and B's detectors, as in this regime the 6-terminal scattering matrix includes only back reflection terms. The presented result is therefore only valid in the limit $\gamma_\text{sf}=0, \gamma_\text{sp}\rightarrow 1$, where the normalization of the particle-number-difference correlator, Eq.~\eqref{eq:correlators}, ensures convergence. Hence, the post-selection brings the disadvantage, that only a fraction of the particle-pairs produced by the SpPS shows non-classical (entangled) correlations.  Interestingly the maximal violation of the CHSH inequality is obtained in a regime, where the fraction of particles reaching the two detectors is vanishingly small. Therefore, in the proposed set-up, one has to weigh between the rate of production of entangled particle pairs and the quality of the entanglement. Similar results are obtained in the proposal of Ref.~[\onlinecite{Lebedev:2004}]. However, the polarization angles leading to maximum violation are slightly different.

%
%
%
%


\section{Conclusions and Discussion}
We have extended the concept of Single-Particle Sources to a system with time reversal symmetry. The proposed device emits pairs of spin-polarized charged particles into the edges of a two-dimensional topological insulator, justifying the name Spin Particles Source. Additionally, the two emitted particles are simultaneously moving in opposite directions. We have suggested a noise measurement that permits to investigate the helical nature of the two electrons in a new phenomenon, called $\mathbb{Z}_2$-\textit{antibunch}. In addition, such a noise measurement can proof the simultaneous emission of two  particles. Finally, we have proposed a set-up allowing for the entanglement of the two emitted particles, manifested in a non vanishing concurrence and a violation of a CHSH inequality. Our scheme is based on post-selection, even if this is a very debate point in the quantum information community.\cite{hannes:2008} It is especially questionable, whether entanglement produced by post-selection is useful, in the sense, that destruction of the entangled particles is required in order produce the entanglement. Hence, the entanglement is not available for further quantum information usage. However, we are convinced that post-selection is preserving the original spirit of the EPR ``Gedankenexperiment".~\cite{EPR} We further believe that our set-up might be useful for quantum cryptography in the spirit of \emph{e.g.} the BB84-protocol.\cite{BB84} Other set-ups for the entanglement of  the edge states of 2D TI using superconductors have been also investigated.~\cite{chen:2012}  

An important remark has to be done regarding the feasibility of the SpPS. It is based on the possibility of separating a QD from the external part via a QPC. As stated in the introduction, at the moment only two 2D TI are known, \emph{i.e.} HgTe/CdHgTe and InAs/GaSb quantum wells. For these materials, no experimental results for QPCs have been reported so far. However, InAs/GaSb contrary to HgTe/CdHgTe has the advantage that the transition to the TI phase can be tuned via electrostatic gating of the material.~\cite{liu:2008,knez:2011} Therefore, this property could be exploit in order to realize quantum constrictions by electrostatic gating. A gate provides a modulation of the electrostatic profile in the heterostructure. In a standard semiconducting mesoscopic device this corresponds to a quantum constriction, however in the case of InAs/GaSb quantum wells this would create a boundary between a topological and a trivial insulator.~\cite{footnote:du}

Even though experimental knowhow on 2D TI is limited so far, we believe that experimental implementation of the SpPS has a twofold interest: First, it can have applications in spintronics, where pure spin-currents are a key tool. Second, it could be a realization of a Bell-test on a chip in a framework different from the IQHE. Thus our proposal  could allow to transfer quantum optics paradigms to solid state systems, complementing the field of the so called \emph{electron quantum optics}.


\acknowledgments 
We acknowledge H.-P.~Breuer, P.~Brouwer, A.~de Martino, P.~Lucignano, L.~Lenz, M.~Rizzi, L.~Sch\"atzle, and J.~Splettstoesser for useful discussions. This work is dedicated to the memory of Markus B\"uttiker. We had the honor to discuss with him about several aspects of the SpPS. The work of DB is supported  by the Excellence Initiative of the German Federal and State Governments and the Alexander von Humboldt Foundation. The work of AI is supported by the Studienstiftung des Deutschen Volkes.

\appendix

\section{The system scattering matrix}\label{app:one}
\begin{figure}
\includegraphics[width=0.9\columnwidth]{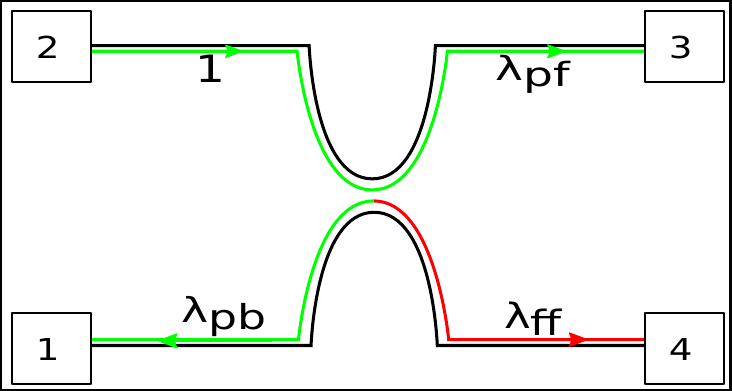}
\caption{ (Color online) The quantum point contact (QPC) connecting the edge-states to the quantum dot. Propagation along the edges is spin polarized, such that on the upper edge spin-up [spin-down] propagates to the right [left] and vice-versa on the lower edge. A constriction in the bulk structure makes the edge-states overlapping what allows for different scattering processes described by the Hamiltonians \eqref{eq:zero} and \eqref{eq:qpc}.\label{FourTerminals}}
\end{figure}

For determining the scattering matrix of the QPC, the dot is enlarged allowing to consider a four-terminal set-up as depicted in Fig.~\ref{FourTerminals}.
The relevant Hamiltonians at the constriction are expressed by
%
%
\begin{subequations}\label{eq:qpc}
\begin{align}
\label{eq:pf_Hamiltonian}
\mathcal{H}_\text{tun}^\text{sp}& =\sum_{\sigma=\uparrow,\downarrow} \int dx \  \nonumber\left[\Gamma_\text{sp}(x) \Psi_{\text{R}\sigma}^\dag(x) \Psi_{\text{L}\sigma}  \right.\\
&\hspace{2cm}+\left.\Gamma_\text{sp}^*(x) \Psi_{\text{L}\sigma}^\dag(x) \Psi_{\text{R}\sigma}(x)\right]\\
\label{eq:ff_Hamiltonian}
\mathcal{H}_\text{tun}^\text{sf}& =\sum_{\alpha=\text{L,R}} \xi_\alpha\int dx \	 \left[\Gamma_\text{sf}(x) \Psi_{\alpha\uparrow}^\dag(x) \Psi_{\alpha\downarrow} \right.\nonumber \\ & \hspace{2cm}+\left.\Gamma_\text{sf}^*(x) \Psi_{\alpha\downarrow}^\dag(x) \Psi_{\alpha\uparrow}(x)\right]
\end{align}
\end{subequations}
%
%
Here $\xi_\alpha=\pm 1$ depending on wether the state is a right--mover ($+$/R) or a left--mover ($-$/L) that accounts for chirality of the edge-states. In Eqs.~\eqref{eq:qpc}, the functions $\Gamma_\text{sp}(x)$ and $\Gamma_\text{sf}(x)$ are space-dependent tunneling amplitude profiles for the spin-preserving backscattering and the spin-flipping forward scattering process, respectively. We assume, that the length-scale $\lambda_{\text{QPC}}$, over which the effects of the QPC are non-negligible is small compared to all other relevant system length scales.\cite{Dolcini:2011} Therefore the potential of the QPC can be assumed to be \emph{point}--like. This allows to write down the tunneling amplitudes as 
%
%
\begin{equation*}
\Gamma_\text{sp/sf}(x)=2 \hbar v_\text{F} \gamma_\text{sp/sf}\delta(x),
\end{equation*}
%
%
with $\gamma_\text{sp/sf}$ being real-valued dimensionless tunneling amplitudes for the QPC. The strength of the coefficients $\gamma_{i}$ can be modulated by changing the length of the QPC channel and  also by side gate potentials that change the strength of the spin-orbit interaction.\cite{Krueckl:2011}

The Heisenberg  equations of motion are obtained by calculating
%
%
\begin{equation}
\text{i}\hbar \partial_t \psi_{\alpha,\sigma}=[\psi_{\alpha,\sigma},\mathcal{H}]
\end{equation}
where $\mathcal{H}=\mathcal{H}_0 + \mathcal{H}_\text{tun}^\text{sp}+\mathcal{H}_\text{tun}^\text{sf}$ is the full Hamiltonian [Eqs.~\eqref{eq:zero} and \eqref{eq:qpc}], $\alpha=$L,R stands for right- and left-movers and $\sigma=\uparrow,\downarrow$ is the spin. Note, that spin and direction of propagation are correlated [Fig.~\ref{FourTerminals}]. They read
%
%
\begin{subequations}\label{eq:Heisenberg}
\begin{align}
\text{i} \partial_t \psi_{\text R , \uparrow}(x)=&-\text{i} v_\text F \partial_x \psi_{\text R , \uparrow}(x)\\ \nonumber &+2 v_\text F \delta(x)[\gamma_\text{sp}\psi_{\text L , \uparrow}(x)+ \gamma_\text{sf}\psi_{\text R , \downarrow}(x)]\\
\text{i} \partial_t \psi_{\text R , \downarrow}(x)=&-\text{i} v_\text F \partial_x \psi_{\text R , \downarrow}(x)\\ \nonumber &+2 v_\text F \delta(x)[\gamma_\text{sp}\psi_{\text L , \downarrow}(x)+ \gamma_\text{sf}^*\psi_{\text R , \uparrow}(x)]\\
\text{i} \partial_t \psi_{\text L , \uparrow}(x)=&+\text{i} v_\text F \partial_x \psi_{\text L , \uparrow}(x)\\ \nonumber &+2 v_\text F \delta(x)[\gamma_\text{sp}^* \psi_{\text R , \uparrow}(x)- \gamma_\text{sf}\psi_{\text L , \downarrow}(x)]\\
\text{i} \partial_t \psi_{\text L , \downarrow}(x)=&+\text{i} v_\text F \partial_x \psi_{\text L , \downarrow}(x)\\ \nonumber &+2 v_\text F \delta(x)[\gamma_\text{sp}^* \psi_{\text R , \downarrow}(x)- \gamma_\text{sf}^*\psi_{\text L , \uparrow}(x)]
\end{align}
\end{subequations}
For a fixed energy, the solutions are found via the ansatz \cite{Dolcini:2011} of plane waves
%
%
\begin{equation}\label{eq:Ansatz}
\begin{aligned}
\psi_{\text R,\sigma}(x)&=&\frac{\text{e}^{-i\frac{E}{\hbar}t}}{\sqrt{hv_\text F}}\begin{cases}
a_{\text R, \sigma} \text{e}^{\text{i} k_E x} & x<0\\
b_{\text R, \sigma} \text{e}^{\text{i} k_E x} & x>0\\
\end{cases}\\
\psi_{\text L,\sigma}(x)&=&\frac{\text{e}^{-i\frac{E}{\hbar}t}}{\sqrt{hv_\text F}}\begin{cases}
b_{\text L, \sigma} \text{e}^{-\text{i} k_E x} & x<0\\
a_{\text L, \sigma} \text{e}^{-\text{i} k_E x} & x>0\\
\end{cases}
\end{aligned}
\end{equation}
with $k_E=\frac{E}{\hbar v_\text F}$.

The scattering matrix of the QPC $\Sigma$ is defined via
%
%
\begin{equation}
\left(b_{\text{L}\uparrow},b_{\text{L}\downarrow},b_{\text{R}\uparrow}, b_{\text{R}\downarrow}\right)^T = \Sigma \cdot \left( a_{\text{R}\downarrow} , a_{\text{R}\uparrow}, a_{\text{L}\downarrow},a_{\text{L}\uparrow}\right)^T
\end{equation}
where $T$ stands for transposition.

Plugging the ansatz~\eqref{eq:Ansatz} into Eqs.~\eqref{eq:Heisenberg} and solving the set of linear equations, yields the result presented in the main text, Eq.~\eqref{QPS:SCM}. 
%
%
%
%

\section{The time-dependent scattering matrix}\label{app:two}

Here, we calculate the full time-dependent scattering matrix of the quantum dot. The potential applied to the gate is assumed to be periodic in time with period $\mathcal{T}$. According to the Floquet theorem, the dot can thus change the energy $E$ of a particle to energies $E_n=E+n\hbar\Omega_0$ with $\Omega_0=2\pi/\mathcal{T}$. The theoretical framework for describing such scattering processes was developed in Ref.~[\onlinecite{Moskalets:2011}] and goes under the name of Floquet scattering matrix.
The states inside the QD are given in the $s_z$ basis by
%
%
\begin{align}
\Psi_{\text{QD}}(y,t)=& \text{e}^{-\text{i}Et/\hbar}\mathcal{Y}(t) \begin{pmatrix}a\left(t+\frac{y}{v_\text{F}}\right)\text{e}^{-\text{i}k_Ey}\\
 b\left(t-\frac{y}{v_\text{F}}\right)\text{e}^{\text{i}k_Ey}\end{pmatrix}
\end{align}
%
%
where $a(t)$ and $b(t)$ are time-dependent amplitudes and $\mathcal{Y}(t)=\operatorname{exp}\left[-\text{i}\frac{e}{\hbar}\int_{-\infty}^t  U(t')dt'\right]$ accounts for the time-dependent gate-voltage applied to the dot. For the moment, consider a spin-up electron with energy $E$ incident from the left lead: The edge-state in the left half ($x>0$) in its most general form can be written as
%
%
\begin{equation}
\Psi^{\text{left}}(t,x)=\begin{pmatrix}
\text{e}^{-\text{i} E t /\hbar + \text{i} k_E x}\\
\displaystyle\sum_{n=-\infty}^\infty \text{e}^{-\text{i} E_n t /\hbar - \text{i} k_n x}\mathcal{S}_{\text{F,22}}(E_n,E)
\end{pmatrix}.
\end{equation}
%
%
The underlying linear dispersion ($k_n=k_E+n\Omega_0/v_\text{F}$) allows to write
%
%
\begin{align}
\Psi^{\text{left}}(t,x)=&\text{e}^{-\text{i} \frac{E t }{\hbar}-\text{i} k_E x}\times \\ &\begin{pmatrix}
\text{e}^{2\text{i} k_E x}\\
\displaystyle\sum_{n=-\infty}^\infty \text{e}^{-\text{i} n\Omega_0 (t+x/v)} \mathcal{S}_{\text{F,22}}(E_n,E)
\end{pmatrix}\nonumber.
\end{align}
%
%
Introducing the time-dependent function
%
%
\begin{equation}
\mathcal{S}^{\text{(in)}}(t,E)=\sum_{n=-\infty}^{\infty} \text{e}^{-\text{i}n\Omega_0 t} \mathcal{S}_F(E_n,E)
\end{equation}
%
%
this reads
%
%
\begin{equation}
\Psi^{\text{left}}(t,x)=\begin{pmatrix}
\text{e}^{-\text{i} E t /\hbar + \text{i} k_E x}\\
\text{e}^{-\text{i} E t /\hbar-\text{i} k_E x} \mathcal{S}_{\text{22}}^{\text{(in)}}(t+\frac{x}{v_\text{F}},E)
\end{pmatrix}.
\end{equation}
%
%
Correspondingly the state in the right half reads
%
%
\begin{equation}
\Psi^{\text{right}}(t,x)=\begin{pmatrix}
\text{e}^{-\text{i} E t /\hbar+\text{i} k_E x} \mathcal{S}_{\text{12}}^{\text{(in)}}(t-\frac{x}{v_\text{F}},E)\\
0
\end{pmatrix}.
\end{equation}
%
%
As the time-origin can always be chosen such that the gate-voltage does not break time-reversal symmetry, we expect $S_{22}(t,E)$ to vanish. This is shown explicitly here.

Consider the loop to have circumference $L$ and the QPC to be located at $x=y=0$. Then the states are relied via the scattering matrix $\Sigma$ of the QPC Eq.~\eqref{QPS:SCM} as follows
%
%
\begin{equation}\label{eq:QPC}
\begin{pmatrix}
\mathcal{S}_{12}^{\text{(in)}} (t,E)\\
\mathcal{S}_{22}^{\text{(in)}} (t,E)\\
\mathcal{Y}(t) a(t+\tau)\text{e}^{-\text{i}k_E L}\\
\mathcal{Y}(t) b(t)
\end{pmatrix}
=
\Sigma
\cdot 
\begin{pmatrix}
0\\
1\\
b(t-\tau)\mathcal{Y}(t)\text{e}^{\text{i}k_E L}\\
a(t)\mathcal{Y}(t)
\end{pmatrix}.
\end{equation}
%
%
This yields the defining equations for $b(t)$ and $a(t)$:
%
%
\begin{subequations}
\begin{align}
b(t)\mathcal{Y}(t)=&\lambda_\text{ff}^* +\lambda_\text{pb} \mathcal{Y}(t) b(t-\tau) \text{e}^{\text{i} k_\text{E} L}\\
a(t+\tau)\mathcal{Y}(t)\text{e}^{-\text{i} k_\text{E} L} =&\lambda_\text{pf} +\lambda_\text{pb} \mathcal{Y}(t) a(t)
\end{align}
\end{subequations}
%
%
Where the classical time for one turn inside the dot $\tau=L/v_\text{F}$ was introduced.
%
%
\begin{figure}[!t]
	\centering
	\includegraphics[width=0.9\columnwidth]{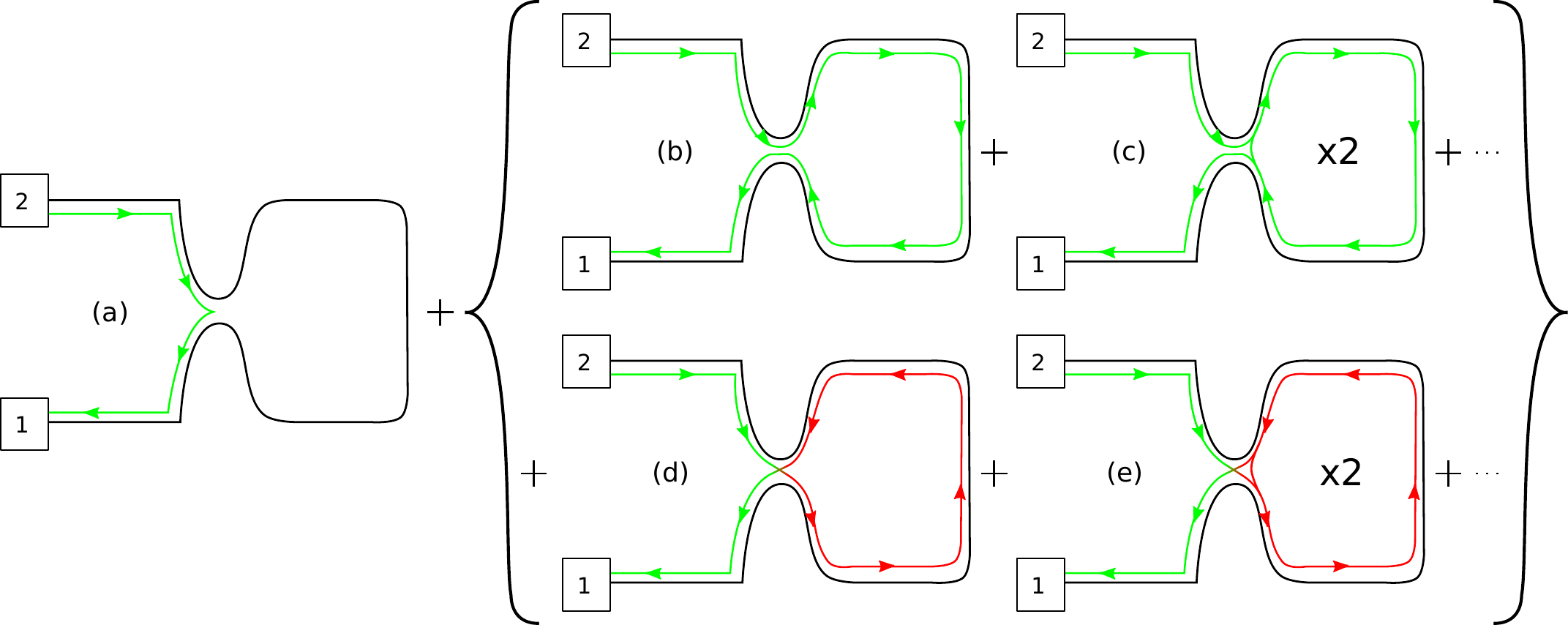}
	\caption{\label{figure:processes} (Color online) In the figure are shown the processes contributing to the time-dependent scattering matrix Eqs.~\eqref{eq:time:dep:s}. The panel (a) shows the back-reflection, proportional to $\lambda_\text{pb}$, panels (b) and (c) (green line) are the forward spin conserving processes proportional to $\lambda_\text{ff}$, and panels (d) and (e) (red line) are the forward spin flipping processes proportional to $\lambda_\text{pf}$. Each time the edge states are arriving to the QPC they get a contribution proportional to $\lambda_\text{pb}$.}
\end{figure}
%
%

\noindent These equations allow for the following solutions
%
%
\begin{subequations}
\begin{align}
b(t)=&\lambda_\text{ff}^* \sum_{q=0}^\infty \lambda_\text{pb}^q \text{e}^{\text{i} q k_E L} \mathcal{Y}^*(t-q\tau)\\
a(t)=&\lambda_\text{pf} \sum_{q=0}^\infty \lambda_\text{pb}^q \text{e}^{\text{i} (q+1) k_E L} \mathcal{Y}^*(t-(q+1)\tau)
\end{align}
\end{subequations}
%
%
Together with equation \eqref{eq:QPC}, the expressions for the scattering matrix elements are found:
%
%
\begin{subequations}\label{eq:time:dep:s}
\begin{align}
\mathcal{S}_{12}^{\text{(in)}}(t,E)=&\lambda_\text{pb} + (|\lambda_\text{ff}|^2 + \lambda_\text{pf}^2)  \\ \nonumber& \times\sum_{q=0}^\infty \lambda_\text{pb}^q \text{e}^{\text{i} (q+1) k_\text{E} L} \mathcal{Y}^*(t-(q+1)\tau)\mathcal{Y}(t)\\
\mathcal{S}_{22}^{\text{(in)}}(t,E)=&\lambda_\text{pf}(\lambda_\text{ff}^* + \lambda_\text{ff}) \\ \nonumber & \times\sum_{q=0}^\infty \lambda_\text{pb}^q \text{e}^{\text{i} (q+1) k_\text{E} L} \mathcal{Y}^*(t-(q+1)\tau)\mathcal{Y}(t)
\end{align}
\end{subequations}
%
%
From this expression, one directly sees that $\mathcal{S}_{22}$ is proportional to 
%
%
\begin{equation}
\lambda_\text{ff}+\lambda_\text{ff}^*=2\text{Re} \left[\lambda_\text{ff}\right]\,.
\end{equation}
%
%
If $\lambda_\text{ff}$ is purely imaginary|as requested by TRS|the diagonal elements of the scattering matrix vanish.\cite{qi:2010} The case of the incident particle being a spin-down electron propagating to the left is done analogously and yields the results presented in the main text. The various processes constituting the scattering matrix element $\mathcal{S}_{12}(E,t)$ are illustrated in Fig.~\ref{figure:processes}. The first term proportional to $\lambda_\text{pb}$ corresponds to back reflection at the QPC Fig.~\ref{figure:processes}(a). The other processes corresponds to turning clockwise [Fig.~\ref{figure:processes}(b) and \ref{figure:processes}(c)] or anti clockwise inside the QD [Fig.~\ref{figure:processes}(d) and \ref{figure:processes}(e)] and  are proportional to $\lambda_\text{pf}$ and to $|\lambda_\text{ff}|^2$, respectively.


\section{The time-dependent current}\label{app:three}
The current operator in lead $\alpha$ is given as\cite{buettiker:1992}
%
%
\begin{equation}
\hat I_\alpha= \frac{e}{h}\iint_0^\infty dE dE' \text{e}^{\text{i} \frac{(E-E')t}{\hbar}} \left[ \hat b(E)\hat b(E') -\hat a(E) \hat a(E')\right]
\end{equation}
%
%
where $a(E)$ [$b(E)$] is an annihilation operator for an electron with energy $E$ incident on [scattered off] the scattering region. In the time-dependent case, the $b(E)$-operators are related to the $a$ operators via the Floquet scattering matrix
%
%
\begin{equation}
b(E)=\sum_{n=-\infty}^\infty S_\text{F} (E,E_n) a(E_n)\,.
\end{equation}
%
%
Due to the symmetries of the considered set-up, the current flowing in lead $1$ is the same as the current flowing in lead $2$. We assume here the case of identical leads, characterized by Fermi-distributions $f_0(E)$. In terms of the time-dependent scattering matrix $\mathcal{S}_{ij}(t,E)$, the current reads then
%
%
\begin{align}
I_1(t)=\frac{e}{h}&\int_{-\infty}^\infty dE \sum_{n=-\infty}^{\infty} \left[f_0(E)-f_0(E_n)\right]\\ \nonumber &\times\int_0^\mathcal{T}\frac{dt^\prime}{\mathcal{T}} \text{e}^{\text{i} n\Omega_0 (t-t^\prime)} S_{12}(t,E)S_{12}^*(t^\prime,E)\, .
\end{align}
%
%
Plugging Eq.~\eqref{eq:scattmat:2} into  this equation yields the result $I_\alpha(t)=I_\alpha^\text{d}+I_\alpha^\text{od}$.


\end{document}